\documentclass[sigconf]{acmart}
\AtBeginDocument{%
  }

\copyrightyear{2026}
\acmYear{2026}
\setcopyright{cc}
\setcctype{by}
\acmConference[MM '26]{Proceedings of the 34th ACM International Conference on Multimedia}{November 10--14, 2026}{Rio de Janeiro, Brazil}
\acmBooktitle{Proceedings of the 34th ACM International Conference on Multimedia (MM '26), November 10--14, 2026, Rio de Janeiro, Brazil}
\acmDOI{10.1145/3767308.3836600}
\acmISBN{979-8-4007-2213-4/2026/11}

\usepackage{tabularx}
\usepackage{pifont}
\usepackage{multirow}
\usepackage{textgreek}
\usepackage[commandnameprefix=always]{changes}
\usepackage{caption}

\usepackage{hyperref}
\usepackage{subfig}
\usepackage{graphicx} 
\usepackage[table]{xcolor}
\usepackage{makecell}

\newcommand\Mansi[1]{\textcolor{blue}{}}
\newcommand\philipp[1]{\textcolor{red}{}}
\newcommand\Dinar[1]{\textcolor{brown}{}}

\newcommand{\dataset}{\textsc{EEGaze-140}}
\newcommand{\method}{\textsc{EEGazeNet}}

\begin{document}



\title{\textsc{EEGaze}: Multimodal Attention Prediction in Visual Search via EEG-Based Saliency Refinement}





\author{Mansi Sharma}
\email{msharma@is.mpg.de}
\affiliation{%
  \institution{Max-Planck-Institute for Intelligent Systems}
  \city{Stuttgart}
  \country{Germany}
}
\affiliation{%
  \institution{Saarland University and DFKI}
  \city{Saarbr\"ucken}
  \country{Germany}
}

\author{Dinar Orazgaliyev}
\email{dior00002@stud.uni-saarland.de}
\affiliation{%
  \institution{Saarland University}
  \city{Saarbr\"ucken}
  \country{Germany}
}

\author{Antonio Kr\"uger}
\email{krueger@dfki.de}
\affiliation{%
  \institution{Saarland University and DFKI}
  \city{Saarbr\"ucken}
  \country{Germany}
}

\author{Philipp M\"uller}
\email{mueller@is.mpg.de}
\affiliation{%
  \institution{Max-Planck-Institute for Intelligent Systems}
  \city{Stuttgart}
  \country{Germany}
}   
\renewcommand{\shortauthors}{Mansi Sharma, Dinar Orazgaliyev, Antonio Krüger, and Philipp Müller}
\begin{abstract}

Visual search is ubiquitous in daily life, from locating keys in a messy apartment to finding a tool in a cluttered workshop or an application icon on a desktop screen.
While visual attention is to a large extent driven by image content, research in psychology indicates that user's cognitive states have a significant influence on attention allocation.
A promising way to access information about cognitive states is via Electroencephalography (EEG) recordings.
Despite this potential, previous works in computational attention modeling have largely disregarded EEG, and related studies in psychology operate with abstract, simplistic stimuli, or even artificially guided search procedures.
%
We bridge this gap with two contributions.
First, we present \dataset, the largest publicly available multimodal dataset of free visual search with realistic stimuli. The dataset contains synchronized EEG and eye-tracking recordings from 36 participants across 140 naturalistic scenes, covering two scenarios: searching for desktop icons and finding tools in cluttered workshop scenes.
Second, we introduce \method, the first method for human attention prediction in visual search that integrates EEG signals.
Our approach consists of a EEG-guided refinement stage that adapts the output of a state-of-the-art saliency prediction method to the user's current cognitive state.
In comprehensive evaluations on \dataset, we show that \method\ consistently outperforms all baselines and ablation conditions.

\end{abstract}




\maketitle

\begin{figure}[ht]
  \centering
  \includegraphics[trim={0cm 0cm 0cm 0cm},clip,width=0.7\columnwidth]{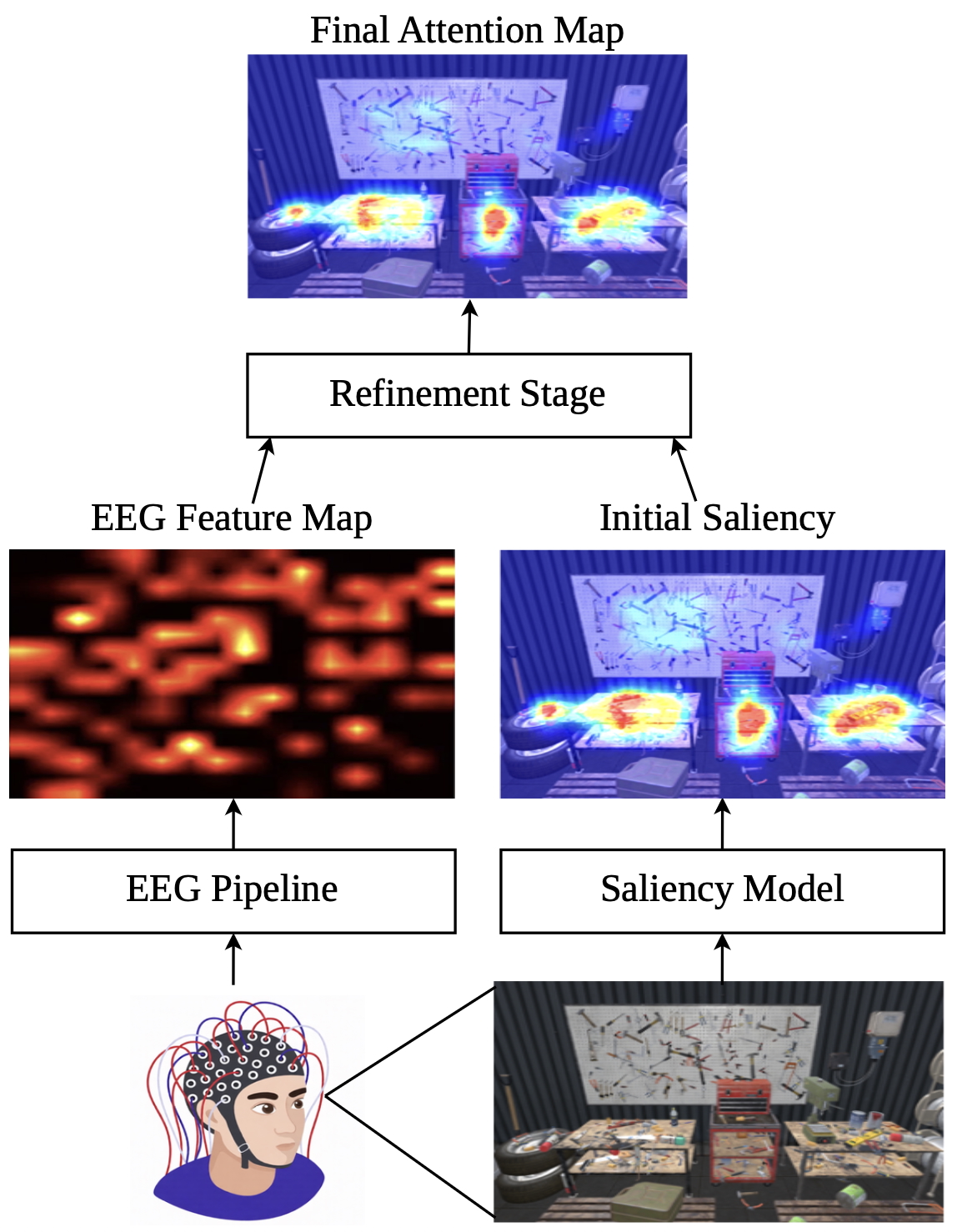}
  \caption{We present \method, the first method to integrate EEG signals into human attention prediction during visual search. The core element of \method\ is a refinement approach that adapts outputs of a state-of-the-art saliency model to the user's cognitive state assessed via EEG prior to the start of the search.}
  \label{fig:teaser}
  \Description[teaser figure]{teaser figure}
\end{figure}

\section{Introduction}
Visual search is a core component of everyday behavior. 
People constantly scan complex environments to find relevant targets, such as keys in a cluttered room, tools in a busy workshop, or application icons on desktop screens. 
Because of this broad relevance, the ability to predict human attention in visual search is important for many interactive systems, including assistive interfaces, adaptive displays, and human-centered AI \cite{hancock2013improving, RN575, RN986, RN987, 10.1145/985692.985780}.
While the scene content plays a decisive role in attention allocation, research in psychology and human factors has shown that human attention is also heavily influenced by cognitive factors including cognitive load \cite{lee2007visual,zagermann2016measuring, sharma2021mental} and mental fatigue \cite{faber2012mental,boksem2005effects}.
As a consequence, effective predictive models of human attention need to integrate image evidence with knowledge concerning the user's cognitive state.


In the last decades, computer vision made impressive progress in modeling those aspects of human attention that are determined by image content.
Recent approaches in the field of saliency prediction utilize CNN, Transformer, or Mamba-based architectures 
\cite{hosseini2025sum, liu2021visual, lou2022transalnet, droste2020unified, aydemir2023tempsal}
in order to predict observer-independent attention distributions during free-viewing.
Some works investigated personalization approaches \cite{10.1145/3655603, xu2018personalized}, or task-specific attention prediction \cite{vozniak2023context, hadnett2019effect}.
However, to the best of our knowledge, no prior works investigated human attention prediction conditioned on general cognitive state beyond simple categorical information, such as task type (e.g. navigation vs. free viewing) or motivational context (e.g. being in a hurry vs. not being in a hurry).
A promising method to access user's cognitive state is via EEG signals.
Research has shown that EEG reflects a wide range of attributes, including alertness \cite{bajaj2020feature}, fatigue \cite{fan2015eeg}, search intent \cite{sharma2023implicit, chiossi2026anticipation}, and cognitive load \cite{vortmann2022multimodal}.
EEG measurements were also shown to be connected to the focus of attention \cite{LIESEFELD2017166, Dolci2023, 10.1093/cercor/10.12.1233}.
This makes EEG a promising additional modality for improving attention prediction beyond uni-modal image-based approaches.
Despite this potential, progress in EEG-based attention modeling remains limited, mainly due to two reasons.
First, there is a lack in sufficiently large, publicly accessible datasets that jointly capture EEG and eye tracking in naturalistic search scenarios.
Second, it is unknown how to effectively integrate EEG signals into human attention prediction pipelines. 
Such integration is challenging, as human attention prediction models are commonly trained on datasets that are magnitudes larger than EEG datasets.

We close this gap with two major complementary contributions.
First, we introduce \dataset\footnote{https://doi.org/10.5281/zenodo.19329128}, a new publicly available 
multimodal dataset of visual search across 140 realistic stimuli. The dataset contains synchronized EEG and eye-tracking recordings from 36 participants covering two practical use cases: searching for application icons on desktop backgrounds and finding tools in cluttered workshop scenes. 
In total, \dataset~contains 5040 search trials, making it the largest naturalistic visual search dataset with EEG and eye-tracking recordings, both in terms of the number of participants, and the number of search trials.
Second, we present \method, the first human attention prediction approach that integrates EEG signals with image evidence.
Central to \method~is a EEG-based refinement stage that transforms the output of a state-of-the-art saliency prediction approach to reflect the user's cognitive state before the start of their search.
%
%
In evaluations on \dataset~
we show that incorporating EEG consistently improves attention prediction across all metrics, both in cross-user and within-user evaluation scenarios.
Overall, our findings highlight the utility of EEG signals for attention modeling and our novel dataset will serve as a crucial foundation for future research on multimodal attention modeling in visual search. 


\begin{figure}
\centering
  \includegraphics[trim={0cm 0cm 0cm 0cm},clip,width=\linewidth]{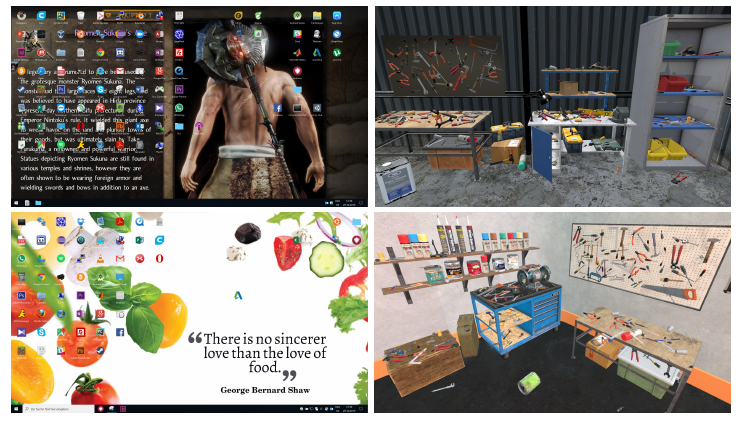}
  \caption{Examples of the desktop and workshop visual scenes we created for our study.}
  \Description[stimuli]{stimuli}
  \label{Stimuli}
\end{figure}

\begin{table*}
\centering
\caption{Datasets for visual search task. Availability shows whether the dataset is publicly available or not. Size is a product of the total number of users and scenes shown to each user. $^\ast$ denotes that only a single type of realistic visual stimuli is used, whereas our dataset comprises of two scene types: workshop and desktop scenes.}
  \label{tab:RW}
  {\begin{tabular}{lcccccc}
    \toprule
    Reference & Available & Users  & Size  & Search & Stimuli \\
    \midrule
    \citet{brouwer2017eeg} & \ding{55}  & 21  &  176 & Guided & Abstract  \\
     \citet{kastrati2021eegeyenet} & \ding{51}  & 356  & 21360  & Guided & Abstract \\
    \citet{kamienkowski2012fixation} & \ding{55}  & 17 & 209 & Free & Abstract
    \\
   \citet{sharma2023implicit} & \ding{51}  & 15   & 1800  & Free & Realistic $^\ast$  \\
    \citet{kaunitz2014looking} & \ding{55}  & 12  & 2160 & Free & Realistic $^\ast$  \\
    \midrule
\dataset & \ding{51} & 36 & 5040 
& Free &
\makecell[l]{\hspace{1cm} Realistic \\(Workshop and Desktop)} \\

 \bottomrule
\end{tabular}}
\end{table*}

\section{Related Work}
\label{sec:RW}
We review prior work on visual attention prediction, discuss EEG-based attention links, and summarize existing EEG and eye-tracking datasets for visual search.

\subsection{Visual Attention Prediction}
Research on human attention prediction has a long history in cognitive science and computer vision.
A major interest in this field of research is saliency prediction, i.e. the task of predicting where humans will look based on the characteristics of an image.
While early work was grounded in human attention theory \cite{koch1987shifts,itti2002model}, more recent approaches employ deep learning architectures.
The initial CNN-based models such as eDN \cite{vig2014large} and DeepGaze I \cite{kummerer2014deep} used pre-trained networks for transfer learning. 
Later models improved performance using deeper architectures and semantic representations, including DeepFix (VGG-16 backbone) \cite{kruthiventi2017deepfix} and DeepGaze II (VGG-19 backbone) \cite{kummerer2016deepgaze}. 
Recent approaches shifted toward Transformer-based and hybrid architectures. TranSalNet combines CNN backbones with Transformer modules to model global dependencies \cite{lou2022transalnet}, while VST adopts fully Transformer-based encoding with multi-level token fusion and task-specific decoding \cite{liu2021visual}. 
Most recently, SUM \cite{hosseini2025sum} introduced a cross-domain unified framework using a U-Net backbone with C-VSS modules and Mamba blocks, achieving state-of-the-art performance across multiple datasets. 

Compared to the large body of work on free-viewing saliency prediction, fewer works tried to incorporate task characteristics into attention prediction models \cite{vozniak2023context, zheng2018task, ramanishka2017top}. 
For example, Zheng et al. \cite{zheng2018task} proposed a framework disentangling task and task-free saliency, demonstrating that task-driven models outperform standard baselines for webpage saliency prediction.
Context-SalNET \cite{vozniak2023context} incorporated task characteristics such as urgency, and different safety preferences into pedestrian attention prediction.
Building on these task-specific approaches, others have further explored attention in visual search task by identifying distracting objects \cite{samiei2022predictingvisualattentiondistraction} and modeling the impact of question-driven saliency in information visualizations \cite{wang2024salchartqa}.
Beyond universal models, Strohm et al. \cite{10.1145/3655603} introduced a Siamese convolutional encoder that leverages implicit gaze embeddings to personalize saliency maps.

Overall, while current models account for task characteristics and user identity, they largely overlook the subject's internal cognitive states, which could serve as a highly informative source for refining saliency predictions.


\subsection{EEG and Visual Attention}

EEG signals have been linked to human attention allocation across a wide variety of settings.
Dolci et al. \cite{Dolci2023} and Hopf et al. \cite{10.1093/cercor/10.12.1233} showed that EEG markers, such as N2pc, P1 and CNV, reveal how the brain builds a topographical visual priority map. 
Extending these foundational EEG markers to more applied context, some studies have examined neural correlates of attention during natural viewing. Van Humbeeck et al. \cite{van2018presaccadic} used co-registered EEG and eye-tracking to analyze pre-saccadic activity, showing that fixating less salient targets requires greater neural effort. 
Farahat et al. \cite{farahat2019convolutional} used CNNs to decode covert attention from EEG and applied attention maps to interpret which neural features influenced model decisions, highlighting how attention modeling can enhance EEG-based decoding.
Khaleghi et al. \cite{khaleghi2022visual} introduced GDN-GAN, which reconstructs attention maps from EEG using Chebyshev graph convolutions to model functional connectivity. 
While this previous research documents the promising potential of EEG in attention prediction, it is limited in two crucial aspects. 
First, to the best of our knowledge, no previous work integrated EEG signals as an additional modality into image-based attention prediction approaches.
Second, previous works on EEG-based attention prediction make use of EEG signals recorded at the same time the user looks at the stimulus.
In contrast, our work studies the integration of EEG signals recorded prior to stimulus presentation into state-of-the-art image-based attention prediction approaches.
In this way, our approach can make meaningful predictions about the expected attention distribution once the user starts looking at the stimulus.



\subsection{Visual Search Datasets}
Datasets of EEG and eye-tracking recordings during visual search are scarce.
Table~\ref{tab:RW} summarizes the existing datasets used in previous studies on visual search. 
Most of these datasets use abstract stimuli, and some even guided search procedures, where user's fixation sequences are explicitly instructed.
Furthermore, most datasets are not publicly available.
The most similar dataset to our work is the one by Sharma et al.~\cite{sharma2023implicit} which was initially recorded to study the classification of navigational versus informational intent.
It features realistic industrial workshop scenes, but is limited in size, with only 1800 trials from 15 users performing 120 search tasks each.
In contrast, \dataset~incorporates two types of scenes (workshops and computer desktops), and has the largest size of all publicly available datasets (36 users, 5040 search trials).


\section{Dataset}


\begin{figure*}
\centering
  \includegraphics[trim={0cm 0cm 0cm 0cm},clip,width=\textwidth]{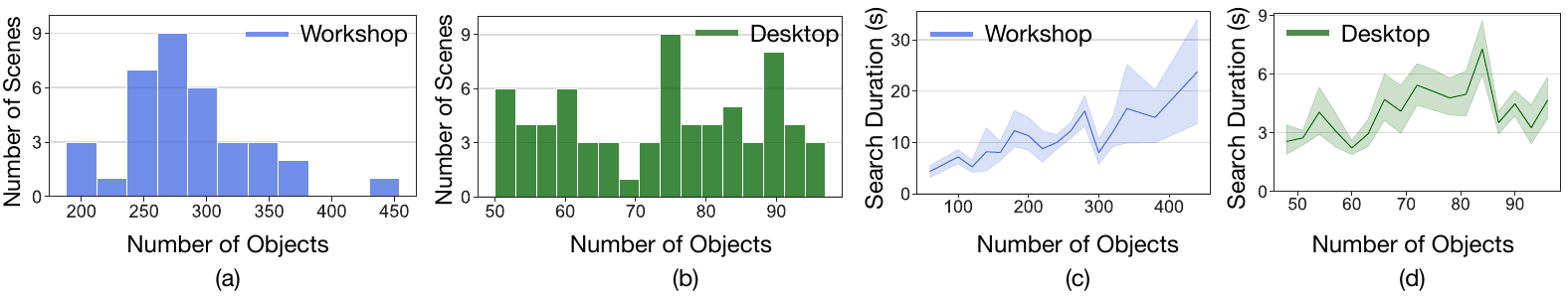}
  \caption{(a), (b) Distribution of number of objects in workshop and desktop scenes, highlighting the controlled complexity of desktops versus the variability and complexity in workshop scenarios. (c), (d) Search duration versus number of objects in workshop and desktop scenes, showing longer search times in workshops compared to desktops search, indicating greater familiarity with desktop setting.}
  \label{Histogram_scenes_obj_search_dur_obj}
  \Description[plots]{plots}
\end{figure*}



In this section, we provide a detailed overview of the data recording process, followed by descriptive statistics and an explanation of the generation of ground truth attention maps. 
\subsection{Data Acquisition}
\paragraph{Participants} We recruited 36 volunteers (17 female and 19 male) aged between 19 and 32 years old ($\mu$~=~23.47, $\sigma$~=~2.69) from our local university. All participants reported normal or corrected-to-normal vision and did not report any known health issues impairing visual perception. None had prior exposure to the study design. The study was approved by our institution’s Ethics and Hygiene Board.

\paragraph{Hardware setup} We presented visual stimuli on a monitor with a resolution of 1920 x 1080 and a screen brightness of 300 cd/m². We recorded EEG signals using the Enobio 20-channel system by Neuroelectrics\footnote{https://www.neuroelectrics.com/solutions/enobio/20}, with a sampling frequency of 500 Hz. We positioned the electrodes according to the 10-20 international electrode placement system. We recorded eye-tracking data using a wireless Tobii Pro Nano device \footnote{https://www.tobii.com/products/discontinued/tobii-pro-nano}, mounted on the lower bezel of the monitor, with a sampling frequency of 60 Hz. The viewing distance from the screen corresponded to approximately $27.54^\circ$ of visual angle. At the beginning of the experiment and after breaks, we calibrated the device for each participant using a 5-point calibration method. 
We recorded and synchronized the eye-tracking data, EEG signals, and experimental events using LabStreamingLayer (LSL) \footnote{https://github.com/labstreaminglayer/App-LabRecorder}.
\paragraph{Stimuli} We constructed 70 realistic scenes using Unity \cite{unity} version 2020.3.41f1 to simulate a wide range of workshop scenes. These scenes included assembly units, manufacturing and production facilities, industrial laboratories, and repair workshops inspired by the research of \cite{sharma2023implicit, sharma2022towards}, 
with implementations based on code made available by the authors.
The design of these virtual scenes aimed to replicate real-world workshop scenes to enhance the ecological validity of the study. 
Each scene was carefully crafted to capture the visual complexity of a typical workplace by arranging machinery, tools, workbenches, and other elements to create varying levels of clutter. We arranged tools in different locations (such as inside cupboards, on the floor, and in unexpected spots) and orientations throughout the scenes. 
Our target stimulus could be one of five tools: Hammer, Pliers, Saw, Screwdriver, or Wrench. On average, each workshop scene contained 250 objects, with a minimum of 64 objects and a maximum of 455, excluding structural elements such as walls, doors, and furniture. 
Additionally, we synthesized 70 desktop backgrounds featuring a broad variety of visual elements, including texts, abstract patterns, and human figures, each with diverse designs, colors, and motifs (see \autoref{Stimuli} for examples). This was based on publicly available code from \cite{muller2022designing}. While designing the desktop scenes, we developed a range of icons and scattered them throughout the background. We utilized a random selection method to determine the number of icons included in each scene. Our target stimulus could include icons like Google Chrome, Instagram, Dropbox, Minecraft, and many others. We assigned a random target icon for each of the 70 scenes. The number of icons in each desktop scene ranged from a minimum of 50 to a maximum of 97, with an average of 73 icons per scene. 
Each target was enclosed by an invisible bounding box used to register mouse clicks and establish the ground truth for target and non-target fixations. The 10-pixel buffer helped account for minor inaccuracies in target detection.

\paragraph{Procedure} We provided participants with a general overview of the study, showcasing the setup and explaining the EEG and eye-tracking sensors to ensure they felt comfortable with the process of tracking their physiological data. Next, we obtained informed consent from the participants and asked them to fill out a demographics questionnaire. We seated participants in a comfortable chair, with a distance to the screen of around 60 cm. The experimenter placed the EEG cap on the participants' heads and filled the electrodes with gel. In total, the preparation took roughly 20-30 minutes. After visually assessing the signal quality, we started data recording, ensuring that electrode impedances remained below 15 k$\Omega$ throughout the experiment. 
A single trial of the experiment comprised three steps. In Step 1, we presented the instructions to begin the study: “Press enter to start”, followed by a 5-second countdown. In Step 2, we displayed the instruction “Search for this item” with the image of the target tool or desktop icon for 5 seconds. Finally, in Step 3, the participant searched for the shown target tool or desktop icon within the scene until they found it. Participants were asked to click on the target using a mouse only after they had located it. This approach minimized distractions from the mouse pointer and reduced the impact of hand movement artifacts on the EEG during the search phase. Participants could skip the search by pressing the 'S' key if they were unable to locate the target. 
We divided the 140 unique scenes into four sessions, including breaks between each session. The first two sessions included workshop scenes, while the last two sessions contained desktop scenes. The sequence of scenes and target objects within each session was randomized. On average, each participant completed the experiment in 120 minutes.



\subsection{Descriptive Statistics}
In total, the dataset includes 5040 visual search trials.
\autoref{Histogram_scenes_obj_search_dur_obj} (a)(b) depicts the distribution of the number of objects in scenes for workshop and desktop scenarios, respectively. 
Workshop scenes generally contain a larger number of objects, with most scenes containing between 250 and 300 objects.
Our desktop scenes on the other hand have generally less objects and a more even distribution from 50 objects per scene to up to 100 objects.
In \autoref{Histogram_scenes_obj_search_dur_obj} (c) (d) we present the relation between the number of scene objects and the average search time.
In workshop scenes, the increase in search duration as the number of objects rises (from 50 to about 450) suggests that a higher number of objects poses a greater challenge to the search task. On the other hand, in desktop scenes, where the number of objects is generally lower (50–90), the average search durations remain below 10 seconds, with a slight increase in time as the object count rises. 
One factor contributing to the pronounced rise of search times in the workshop setting might be the greater likelihood of occlusions when more objects are in the scene.
In desktop scenes, occlusions between objects were not possible.

\begin{figure}
\centering
  \includegraphics[trim={0cm 0cm 0cm 0cm},clip,width=\linewidth]{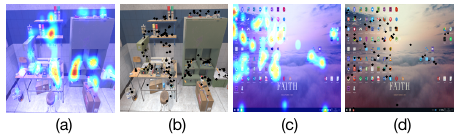}
  \caption{Example saliency and binary fixation maps for Workshop (a–b) and Desktop (c–d) stimuli.}
  \Description[saliency and binary fixation maps]{saliency and binary fixation maps}
  \label{Saliency_Fix_maps}
\end{figure}

\subsection{Attention Map Generation}

To generate spatial attention maps, raw gaze data is first preprocessed and then transformed into representations that capture users’ visual focus during search tasks. Since fixations represent periods during which humans extract important visual information \cite{borji2015eyes, kaunitz2014looking}, we identify fixations using a velocity-threshold (I-VT) based approach inspired by \cite{olsen2012tobii}, following a multi-step preprocessing pipeline \cite{trabulsi2021optimizing}.
The preprocessing pipeline includes several steps. First, \textit{gap fill-in} is applied to handle short missing segments caused by blinks or tracking loss. Next, \textit{eye selection} averages gaze coordinates from both eyes. We then perform \textit{noise reduction} using a moving median filter to smooth the signal. A \textit{velocity calculator} computes gaze velocity for each sample, which is used by the \textit{I-VT classifier} to distinguish fixations from saccades. To improve robustness, we \textit{merge adjacent fixations} that are likely split due to noise, and finally \textit{discard short fixations} below 60~ms.
The resulting fixation points are then converted into spatial attention representations. Fixation locations are accumulated into a binary fixation map where each fixation contributes equally. This map is smoothed using a Gaussian filter ($\sigma = 40$ pixels), following the MIT300 benchmark \cite{mit-tuebingen-saliency-benchmark}, to generate a continuous saliency map that approximates the spatial distribution of visual attention. The resulting maps are normalized to the [0,1] range and stored as grayscale images for each participant. Figure~\ref{Saliency_Fix_maps} shows an example saliency and corresponding binary fixation maps.

\begin{figure}
\centering
\includegraphics[trim={0cm 0cm 0cm 0cm},clip,width=\linewidth]{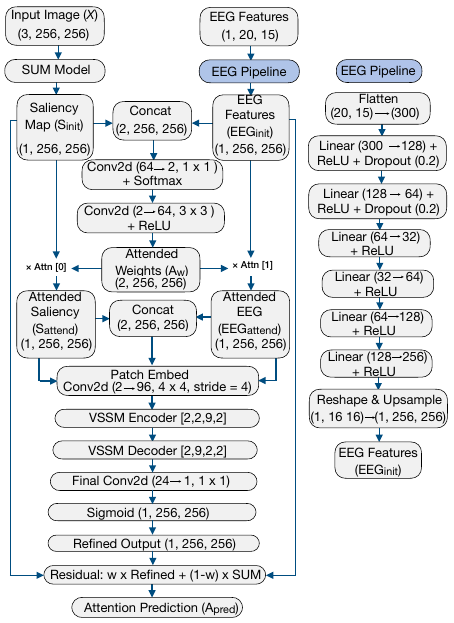}
\caption{Our EEG-based saliency refinement framework generates an initial saliency map with SUM.
EEG features are spatially projected via an MLP pipeline, fused with the SUM output through a channel-wise attention mechanism, refined with a VSSM network, and combined with the initial prediction through residual fusion to produce the final map.}
  \label{our_model}
  \Description[Outline]{res}
\end{figure}

\section{Method}

We propose \method, the first approach that makes use of EEG signals to predict human attention distributions in realistic visual search. Our approach uses an EEG-guided refinement stage to adapt a pretrained saliency model, enabling attention maps sensitive to the user's cognitive state from small EEG datasets.

\subsection{\method~Architecture}
The architecture of \method~is illustrated in \autoref{our_model}.
The input image $X \in \mathbb{R}^{H \times W \times 3}$ is passed through the SUM model, a state-of-the art image-based saliency prediction approach \cite{hosseini2025sum} to produce initial saliency maps $S_{\text{init}}\in \mathbb{R}^{H \times W}$.
\method~subsequently transforms $S_{\text{init}}$ via a EEG-guided refinement network.
To this end, EEG features are processed via a lightweight EEG pipeline, to obtain an initial EEG representation $EEG_{\text{init}}\in \mathbb{R}^{H \times W}$ of same dimensionality as $S_{\text{init}}$. 
$EEG_{\text{init}}$ is obtained as follows.
The input EEG features (see \autoref{sec:eeg_features}), are flattened and passed through a series of fully connected layers with non-linear activations and dropout. 
The resulting latent representation is then projected into a spatial feature map, reshaped to a low-resolution grid, and upsampled via bilinear interpolation to match the spatial resolution of the image ($H \times W$). 
In this way, we obtain an EEG-derived spatial prior aligned with the visual domain.
Subsequently the initial saliency map, $S_{\text{init}}$ and the EEG feature map $EEG_{\text{init}}$ are treated as complementary spatial modalities and concatenated along the channel dimension to form a two-channel representation. 
A spatial attention module with weights $A_w$ is applied to this representation to adaptively weight the contribution of each modality. 
Concretely, convolutional layers followed by a softmax activation generate modality-specific attention weights, i.e., \(\times\ \text{Attn}[0]\) for the SUM modality and \(\times\ \text{Attn}[1]\) for the EEG modality, which are used to produce attended saliency, $S_{\text{attend}}$, and attended EEG feature maps, $EEG_{\text{attend}}$.
These attended features are concatenated and passed through a patch embedding layer, which reduces spatial resolution while increasing the channel dimensionality, and subsequently processed by a hierarchical encoder–decoder architecture based on a Visual State Space Model (VSSM) backbone. The encoder captures multi-scale contextual dependencies, while the decoder reconstructs high-resolution spatial details.
A final convolutional layer followed by a sigmoid activation produces the predicted attention map. Additionally, a residual connection is used to combine the refined prediction with the original SUM output through a learnable weighting parameter, allowing the model to balance between the initial visual prior and the multi-modal prediction, resulting in the final attention prediction $A_{\text{pred}}$.

\subsection{Loss Function}
Our proposed \method \ model utilizes a multi-component loss function designed to optimize saliency predictions while incorporating EEG data, as inspired by \cite{hosseini2025sum, lou2022transalnet, aydemir2023tempsal, droste2020unified} in visual attention modeling. The loss function integrates standard attention prediction metrics adopted from the SUM model.
The composite loss function is formulated as:
\begin{equation}
   \text{Loss} = L_1 + 7 \cdot L_2
\end{equation}
$L_1$ and $L_2$ are adopted directly from the SUM model \cite{hosseini2025sum}, 
where, 
\begin{equation}
   L_1 = -2 \cdot \text{CC} + 10 \cdot \text{KL} - 1 \cdot \text{SIM} - 1 \cdot \text{NSS}
\end{equation}

This component combines standard attention prediction metrics with negative weights to minimize dissimilarity (KL divergence) and maximize similarity (correlation coefficient, similarity, and normalized scanpath saliency). The KL divergence measures distribution dissimilarity between predicted and ground truth attention maps, while CC assesses linear correlation between predictions and ground truth. SIM evaluates overlap between predicted and actual attention maps $A_{\text{GT}}$, and NSS measures correlation between normalized predictions and fixation points.
\begin{equation}
    L_2 = \text{MSE}(A_{\text{pred}},\ A_{\text{GT}})
\end{equation}
This component directly penalizes pixel-wise differences between prediction and ground truth via mean squared error.

\subsection{Implementation Details} Our attention prediction model is implemented using the PyTorch framework and trained on NVIDIA A100 GPUs with 40 GB memory. 
Training proceeds in two sequential stages. In the first stage, a base SUM model is trained to predict attention maps solely from images.
This model is trained on 30 epochs with early stopping after 4 epochs of no improvement, using the Adam optimizer with a learning rate of $1 \times 10^{-4}$ and a step learning rate scheduler that reduces the learning rate by a factor of 0.1 every 4 epochs. The batch size is set to 16 for the SUM model training. All input images and attention maps are resized to a resolution of 256×256 pixels and normalized using ImageNet statistics (mean=[0.485, 0.456, 0.406], std=[0.229, 0.224, 0.225]). 
In the second stage, pretrained SUM weights initialize the EEG-integrated refinement model, which enhances attention predictions by incorporating temporal and neural activity features. The batch size is set to 8 due to additional EEG input.
\subsection{EEG Features}
\label{sec:eeg_features}

\paragraph{Preprocessing}  We cleaned the EEG data by applying high-pass filters with a cutoff frequency of 1 Hz, a notch filter between 48 Hz and 52 Hz, and a low-pass filter with a cutoff frequency of 40 Hz \cite{sharma2022towards}. Afterward, we removed any bad channels and interpolated them using the spherical interpolation method. We set the threshold for rejecting channel correlation at 0.8 \cite{delorme2004eeglab, GramfortEtAl2013a}. Next, we referenced all channels to a common average \cite{ludwig2009using}. To reduce correlation between electrodes, we conducted independent component analysis using the Second Order Blind Identification (SOBI) algorithm \cite{delorme2004eeglab}, followed by the automatic rejection of components identified as muscle, heart, and eye artifacts at a 95\% threshold. 

\paragraph{Feature Extraction}

\begin{table*}[t]
\centering
\small
\begin{tabular*}{\textwidth}{@{\extracolsep{\fill}} lllccccc}
\toprule
\textbf{Stimuli}  & & \textbf{Model} 
& \textbf{KL $\downarrow$} 
& \textbf{SIM $\uparrow$} 
& \textbf{NSS $\uparrow$} 
& \textbf{CC $\uparrow$} 
& \textbf{AUC $\uparrow$} \\
\midrule

\multirow{4}{*}{Desktop}
& & SUM (No Finetuning) 
& 2.094 $\pm$ 0.100 & 0.240 $\pm$ 0.013 & 0.601 $\pm$ 0.051 & 0.165 $\pm$ 0.010 & 0.745 $\pm$ 0.021 \\
& & SUM (Finetuning) 
& 1.799 $\pm$ 0.110 & 0.323 $\pm$ 0.018 & 1.128 $\pm$ 0.095 & 0.322 $\pm$ 0.020 & 0.819 $\pm$ 0.013 \\
& & \method \ (SUM Frozen) 
& 1.856 $\pm$ 0.142 & 0.303 $\pm$ 0.022 & 1.050 $\pm$ 0.137 & 0.302 $\pm$ 0.042 & 0.797 $\pm$ 0.020 \\
& & \method \ (Randomized EEG Descriptive Features) 
& 2.031 $\pm$ 0.116 & 0.282 $\pm$ 0.020 & 1.057 $\pm$ 0.115 & 0.266 $\pm$ 0.027 & 0.818 $\pm$ 0.015 \\
& & \method \  (Randomized EEG Bandpower Ratios) 
& 2.007 $\pm$ 0.115 & 0.282 $\pm$ 0.020 & 1.017 $\pm$ 0.115 & 0.262 $\pm$ 0.027 & 0.816 $\pm$ 0.015 \\
& & \method \ (EEG Bandpower Ratios) 
& 1.766 $\pm$ 0.115 & 0.318 $\pm$ 0.020 & \textbf{1.203 $\pm$ 0.101} & 0.334 $\pm$ 0.025 & 0.829 $\pm$ 0.013 \\
& & \method \ (EEG Descriptive Features)
& \textbf{1.603 $\pm$ 0.108} & \textbf{0.355 $\pm$ 0.021} & 1.150 $\pm$ 0.106 & \textbf{0.359 $\pm$ 0.029} & \textbf{0.831 $\pm$ 0.014} \\

\midrule

\multirow{4}{*}{Workshop}
&
& SUM (No Finetuning) 
& 1.547 $\pm$ 0.111 & 0.339 $\pm$ 0.020 & 0.703 $\pm$ 0.084 & 0.250 $\pm$ 0.028 & 0.751 $\pm$ 0.013 \\
& & SUM (Finetuning) 
& 1.122 $\pm$ 0.102 & 0.462 $\pm$ 0.022 & 1.273 $\pm$ 0.073 & 0.464 $\pm$ 0.021 & 0.815 $\pm$ 0.010 \\
& & \method \ (SUM Frozen) 
& 1.396 $\pm$ 0.108 & 0.387 $\pm$ 0.023 & 0.961 $\pm$ 0.085 & 0.340 $\pm$ 0.029 & 0.769 $\pm$ 0.013 \\
& & \method \ (Randomized EEG Descriptive Features)
& 1.335 $\pm$ 0.106 & 0.420 $\pm$ 0.023 & 1.423 $\pm$ 0.091 & 0.438 $\pm$ 0.023 & 0.832 $\pm$ 0.011 \\
& & \method \ (Randomized EEG Bandpower Ratios) 
& 1.331 $\pm$ 0.106 & 0.420 $\pm$ 0.023 & 1.422 $\pm$ 0.099 & 0.437 $\pm$ 0.025 & 0.832 $\pm$ 0.011 \\
& & \method \ (EEG Bandpower Ratios) 
& 0.954 $\pm$ 0.105 & 0.503 $\pm$ 0.022 & 1.432 $\pm$ 0.098 & 0.515 $\pm$ 0.024 & 0.837 $\pm$ 0.010 \\
& & \method \ (EEG Descriptive Features)
& \textbf{0.945 $\pm$ 0.104} & \textbf{0.504 $\pm$ 0.023} & \textbf{1.434 $\pm$ 0.093} & \textbf{0.516 $\pm$ 0.023} & \textbf{0.838 $\pm$ 0.011} \\
\bottomrule
\end{tabular*}
\caption{Performance comparison on desktop and workshop stimuli for cross-user settings using previous EEG signals. Results are reported as mean $\pm$ 95\% confidence intervals across users. Arrows indicate whether higher ($\uparrow$) or lower ($\downarrow$) values are better.}
\label{tab:cross_user_res}
\end{table*}

EEG features were extracted from 20 channels for all trials. The feature set is inspired by commonly used EEG representations combining spectral, statistical, and complexity-based measures in prior research \cite{fan2015eeg, bajaj2020feature, sharma2023implicit}.
Signals were first normalized using the mean and standard deviation computed from the training set. 
For each channel, Welch’s method was used to compute PSD and relative band powers for delta (0.5–4 Hz), theta (4–7 Hz), alpha (8–13 Hz), beta (13–30 Hz), and gamma (30–40 Hz) bands. Additional features included Hjorth parameters (mobility and complexity), fractal dimensions (PFD, HFD, DFA), shape descriptors (skewness, kurtosis), and amplitude characteristics (min, max, standard deviation), resulting in a total of 15 features, referred to as EEG Descriptive features. 
To evaluate the contribution of EEG features, we compute EEG band-power ratio features that have been commonly associated with fatigue and attentional states\cite{fan2015eeg}. We calculate PSD for each channel using Welch’s method and four ratios were derived: $(\theta+\alpha)/\beta$, $\alpha/\beta$, $(\theta+\alpha)/(\alpha+\beta)$, and $\theta/\beta$, referred to as Bandpower Ratios.
All features were standardized using a scaler fitted on the training set while preserving the channel-wise structure. 
To allow for the prediction of a user's attention distribution in the upcoming search trial, EEG features are always extracted from the preceding trial.




\section{Evaluation}
\subsection{Evaluation Metrics}
To evaluate the quality of the predicted attention maps, we use two categories of metrics: location-based and distribution-based, as outlined in \cite{hosseini2025sum, bylinskii2018different}. 
Location-based metrics, such as Normalized Scanpath Saliency (NSS) and Area Under the ROC Curve (AUC), compare predictions against binary fixation maps, emphasizing how well the model identifies key fixation points. 
In contrast, distribution-based metrics, including the Correlation Coefficient (CC), Similarity (SIM), and Kullback-Leibler Divergence (KLD), assess the alignment between predicted and ground-truth attention distributions using continuous grayscale maps. For all metrics, higher scores indicate better performance, except for KLD, where lower values reflect greater accuracy.

\subsection{Evaluation Scenarios}
We explore two different evaluation setups to understand the utility of EEG in attention prediction during visual search. 
In the cross-user setting, train/validation/test splits are based on users and images, testing on unseen users and images. In the within-user setting, each user is trained and evaluated separately on different images. Models are evaluated separately for workshop and desktop scenes.
All models are evaluated via 6-fold cross-validation, with an 80/10/10 train, validation, and test split. 
We report average performance with 95\% confidence intervals computed across users.

\begin{table*}[t]
\centering
\small
\begin{tabular*}{\textwidth}{@{\extracolsep{\fill}} l l c c c c c}
\toprule
\textbf{Stimuli} & \textbf{Model} 
& \textbf{KL $\downarrow$} 
& \textbf{SIM $\uparrow$} 
& \textbf{NSS $\uparrow$} 
& \textbf{CC $\uparrow$} 
& \textbf{AUC $\uparrow$} \\
\midrule

\multirow{4}{*}{Desktop}
& SUM (No Finetuning) 
& 2.201 $\pm$ 0.064 & 0.221 $\pm$ 0.010 & 0.636 $\pm$ 0.048 & 0.157 $\pm$ 0.011 & 0.741 $\pm$ 0.007 \\
& SUM (Finetuning) 
& 2.127 $\pm$ 0.065 & 0.217 $\pm$ 0.011 & 0.921 $\pm$ 0.036 & 0.247 $\pm$ 0.012 & 0.761 $\pm$ 0.007 \\
& \method \ (SUM Frozen) 
& 1.965 $\pm$ 0.062 & 0.265 $\pm$ 0.012 & 1.009 $\pm$ 0.051 & 0.271 $\pm$ 0.014 & 0.784 $\pm$ 0.008 \\
& \method \ (Randomized EEG Descriptive Features) 
& 1.990 $\pm$ 0.066 & 0.250 $\pm$ 0.013 & 0.978 $\pm$ 0.044 & 0.270 $\pm$ 0.014 & 0.769 $\pm$ 0.007 \\
& \method \ (Randomized EEG Bandpower Ratios) 
& 1.993 $\pm$ 0.067 & 0.247 $\pm$ 0.013 & 0.966 $\pm$ 0.041 & 0.266 $\pm$ 0.015 & 0.764 $\pm$ 0.007 \\
& \method \ (EEG Bandpower Ratios) 
& 1.875 $\pm$ 0.062 & 0.280 $\pm$ 0.012 & 1.197 $\pm$ 0.041 & 0.328 $\pm$ 0.013 & \textbf{0.811 $\pm$ 0.007} \\
& \method \ (EEG Descriptive Features)
& \textbf{1.867 $\pm$ 0.065} & \textbf{0.281 $\pm$ 0.013} & \textbf{1.202 $\pm$ 0.041} & \textbf{0.330 $\pm$ 0.014} & \textbf{0.811 $\pm$ 0.007} \\

\midrule

\multirow{4}{*}{Workshop}
& SUM (No Finetuning) 
& 1.435 $\pm$ 0.052 & 0.371 $\pm$ 0.010 & 0.864 $\pm$ 0.037 & 0.304 $\pm$ 0.014 & 0.768 $\pm$ 0.005 \\
& SUM (Finetuning) 
& 1.296 $\pm$ 0.047 & 0.390 $\pm$ 0.012 & 1.163 $\pm$ 0.031 & 0.408 $\pm$ 0.013 & 0.794 $\pm$ 0.005 \\
& \method \ (SUM Frozen) 
& 1.304 $\pm$ 0.048 & 0.405 $\pm$ 0.011 & 1.039 $\pm$ 0.026 & 0.379 $\pm$ 0.012 & 0.777 $\pm$ 0.005 \\
& \method \ (Randomized EEG Descriptive Features) 
& 1.262 $\pm$ 0.045 & 0.422 $\pm$ 0.011 & 1.158 $\pm$ 0.027 & 0.413 $\pm$ 0.012 & 0.785 $\pm$ 0.004 \\
& \method \ (Randomized EEG Bandpower Ratios)
& 1.262 $\pm$ 0.044 & 0.422 $\pm$ 0.011 & 1.158 $\pm$ 0.028 & 0.413 $\pm$ 0.012 & 0.786 $\pm$ 0.004 \\
& \method \ (EEG Bandpower Ratios) 
& 1.186 $\pm$ 0.045 & 0.434 $\pm$ 0.011 & 1.278 $\pm$ 0.028 & 0.453 $\pm$ 0.012 & 0.818 $\pm$ 0.004 \\
& \method \ (EEG Descriptive Features)
& \textbf{1.181 $\pm$ 0.045} & \textbf{0.435 $\pm$ 0.011} & \textbf{1.284 $\pm$ 0.028} & \textbf{0.455 $\pm$ 0.012} & \textbf{0.819 $\pm$ 0.004} \\

\bottomrule
\end{tabular*}
\caption{Performance comparison on desktop and workshop stimuli for within-user settings using previous EEG signals. Results are reported as mean $\pm$ 95\% confidence intervals across users. Arrows indicate whether higher ($\uparrow$) or lower ($\downarrow$) values are better.}
\label{tab:within_user_res}
\end{table*}

\subsection{Cross-User Results}

\begin{figure*}
\centering
  \includegraphics[trim={0cm 0cm 0cm 0cm},clip,width=\textwidth]{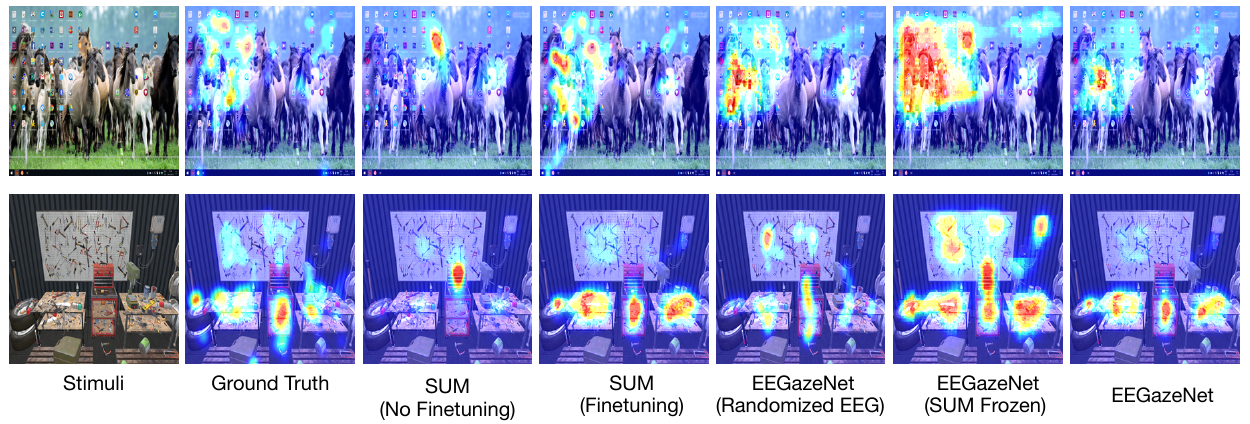}
  \caption{Cross-user attention map visualizations for desktop (first row) and workshop (second row) examples.}
  \label{quali_res}
  \Description[Outline quant]{res quant}
\end{figure*}

\autoref{tab:cross_user_res} summarizes the results of our cross-user experiments.
Most importantly, \method~with EEG Descriptive and Bandpower Ratios features consistently outperforms all baseline and ablation conditions across all metrics, highlighting the effectiveness of EEG integration.
The worst performing approach is SUM without finetuning, highlighting the importance of adapting pre-trained saliency prediction models to the specific target domain.
When comparing \method~to the finetuned SUM baseline, we can observe that the advantage of our approach is slightly more pronounced for workshop scenes compared to desktop scenes.
To further analyze \method, we evaluated three ablation conditions. We first permuted EEG features across trials (\textit{Randomized EEG Descriptive Features and Randomized EEG Bandpower Ratios}) to remove informative EEG signals while keeping the model size unchanged. \method~outperformed this condition across all metrics, confirming the contribution of EEG signals. We furthermore introduce an ablation condition in which we do not fine-tune the SUM model during training (\textit{SUM Frozen}), which performed worse than the full approach, highlighting the importance of joint training, but still outperformed the original SUM model, demonstrating the effectiveness of our refinement approach even without fine-tuning the saliency model.
Participant-level paired Wilcoxon signed-rank tests (N = 36) with Bonferroni correction showed significant differences between \method \ and the fine-tuned SUM baseline across all five metrics (corrected p < 0.05).
In addition to these quantitative results, we present qualitative examples for \method, as well as all baselines and ablation conditions in \autoref{quali_res}.


\subsection{Within-User Results}
We present within-user evaluation results in \autoref{tab:within_user_res}.
The general pattern is in line with the cross-user evaluation.
\method~consistently outperforms all baselines and ablation conditions across all metrics.
A clear difference relative to the cross-user evaluation is visible when comparing the finetuned SUM version to the non-finetuned version: the advantage of fine-tuning is much smaller for the within-user evaluations.
At the same time, the improvements of \method~over both the finetuned and the non-finetuned versions of SUM remain substantial.
For illustration, consider the results for desktop stimuli and the KL metric.
The improvement of fine-tuning the SUM model is 0.295 in the cross-user scenario, and only 0.074 in the within-user scenario.
In contrast, when comparing \method~to the non-fine-tuned SUM model, we observe an improvement of 0.491 in the cross-user case, and an improvement of 0.334 in the within-user scenario. Participant-level paired Wilcoxon signed-rank tests (N = 36) with Bonferroni correction across five metrics showed that \method \ significantly outperformed the fine-tuned SUM baseline across all metrics (corrected p < 0.001).
Finally, when comparing \method~to the fine-tuned SUM model, we observe an improvement of 0.196 in the cross-user scenario, and an even larger improvement of 0.271 in the within-user scenario.

\section{Discussion}

\subsection{Achieved Performance}
Our evaluations showed that \method~outperformed all baselines and ablation conditions across both cross-user and within-user scenarios.
This highlights the effectiveness of EEG integration for attention prediction and our EEG-based saliency refinement approach.
One interesting observation was that, although fine-tuning SUM provided less benefit in the within-user setting, \method~maintained a strong advantage over both fine-tuned and non-fine-tuned SUM versions.
One crucial difference between cross-user and within-user scenarios is the limited training data in the within-user setting, with only one user's attention distribution per image compared to 36 in the cross-user setting.
Another property of the within-user scenario is that all EEG data are recorded from a single person - a decisive advantage given the challenges often associated with EEG in cross-user settings.
The strong performance of \method~in the within-user setting points toward the particular utility of EEG signals in such data-scarce, single-user scenarios.

\subsection{Applications}
Predicting attention during visual search has significant application potential in assistive user interfaces.
If an assistive system knows about the location of a search target, but the user's attention is not predicted to fall on it, the system can support the user by highlighting the respective location, or by guiding the user via verbal instructions.
The more general principle of EEG-based saliency map refinement can also be applied to visual attention prediction outside of search contexts, in order to enable attention-aware, anticipatory human-machine interaction.
In addition to our proposed method, our new publicly available dataset can be used for several additional research directions beyond the one presented in this work.
These include scanpath prediction \cite{xue2025few}, or target fixation detection \cite{brouwer2017eeg, sharma2024distinguishing}.


\subsection{Limitations and Future Work}
While our work is an important first step towards multi-modal attention prediction approaches that integrate EEG signals, several limitations remain.
First, while we make use of highly realistic image stimuli, visual search behavior in real-world environments can pose novel challenges.
One such challenge is the introduction of movement artifacts in EEG signals which need to be treated with dedicated approaches.
Our dataset is furthermore restricted to a particular demographic: relatively young people recruited from a university campus.
Finally, while we covered two highly relevant visual search scenarios, visual search takes place in many more daily situations.
To address these limitations, future research should investigate EEG-based attention prediction in real-world settings, diverse user populations, and broader scenarios.

\section{Conclusion}

In this work, we introduced \dataset, the largest dataset of EEG and eye-tracking recordings during realistic visual search.
We furthermore presented \method, the first approach for attention prediction during visual search informed about a user's cognitive state via integration of EEG signals measured prior to the start of the search.
In comprehensive evaluations on \dataset, we show that our approach consistently outperforms all baselines and ablation conditions.
While \method\ is an important initial step towards EEG-informed attention prediction, our new publicly available dataset will serve as a valuable resource for future research.

\bibliographystyle{ACM-Reference-Format}
\bibliography{sample-base}


\begin{thebibliography}{57}


\ifx \showCODEN    \undefined \def \showCODEN     #1{\unskip}     \fi
\ifx \showISBNx    \undefined \def \showISBNx     #1{\unskip}     \fi
\ifx \showISBNxiii \undefined \def \showISBNxiii  #1{\unskip}     \fi
\ifx \showISSN     \undefined \def \showISSN      #1{\unskip}     \fi
\ifx \showLCCN     \undefined \def \showLCCN      #1{\unskip}     \fi
\ifx \shownote     \undefined \def \shownote      #1{#1}          \fi
\ifx \showarticletitle \undefined \def \showarticletitle #1{#1}   \fi
\ifx \showURL      \undefined \def \showURL       {\relax}        \fi
\providecommand\bibfield[2]{#2}
\providecommand\bibinfo[2]{#2}
\providecommand\natexlab[1]{#1}
\providecommand\showeprint[2][]{arXiv:#2}

\bibitem[Aydemir et~al\mbox{.}(2023)]%
        {aydemir2023tempsal}
\bibfield{author}{\bibinfo{person}{Bahar Aydemir}, \bibinfo{person}{Ludo Hoffstetter}, \bibinfo{person}{Tong Zhang}, \bibinfo{person}{Mathieu Salzmann}, {and} \bibinfo{person}{Sabine S{\"u}sstrunk}.} \bibinfo{year}{2023}\natexlab{}.
\newblock \showarticletitle{TempSAL-uncovering temporal information for deep saliency prediction}. In \bibinfo{booktitle}{\emph{Proceedings of the IEEE/CVF Conference on Computer Vision and Pattern Recognition}}. \bibinfo{pages}{6461--6470}.
\newblock


\bibitem[Bajaj et~al\mbox{.}(2020)]%
        {bajaj2020feature}
\bibfield{author}{\bibinfo{person}{Varun Bajaj}, \bibinfo{person}{Sachin Taran}, \bibinfo{person}{Smith~K Khare}, {and} \bibinfo{person}{Abdulkadir Sengur}.} \bibinfo{year}{2020}\natexlab{}.
\newblock \showarticletitle{Feature extraction method for classification of alertness and drowsiness states EEG signals}.
\newblock \bibinfo{journal}{\emph{Applied Acoustics}}  \bibinfo{volume}{163} (\bibinfo{year}{2020}), \bibinfo{pages}{107224}.
\newblock


\bibitem[Boksem et~al\mbox{.}(2005)]%
        {boksem2005effects}
\bibfield{author}{\bibinfo{person}{Maarten~AS Boksem}, \bibinfo{person}{Theo~F Meijman}, {and} \bibinfo{person}{Monicque~M Lorist}.} \bibinfo{year}{2005}\natexlab{}.
\newblock \showarticletitle{Effects of mental fatigue on attention: an ERP study}.
\newblock \bibinfo{journal}{\emph{Cognitive brain research}} \bibinfo{volume}{25}, \bibinfo{number}{1} (\bibinfo{year}{2005}), \bibinfo{pages}{107--116}.
\newblock


\bibitem[Borji et~al\mbox{.}(2015)]%
        {borji2015eyes}
\bibfield{author}{\bibinfo{person}{Ali Borji}, \bibinfo{person}{Andreas Lennartz}, {and} \bibinfo{person}{Marc Pomplun}.} \bibinfo{year}{2015}\natexlab{}.
\newblock \showarticletitle{What do eyes reveal about the mind?: Algorithmic inference of search targets from fixations}.
\newblock \bibinfo{journal}{\emph{Neurocomputing}}  \bibinfo{volume}{149} (\bibinfo{year}{2015}), \bibinfo{pages}{788--799}.
\newblock


\bibitem[Brouwer et~al\mbox{.}(2017)]%
        {brouwer2017eeg}
\bibfield{author}{\bibinfo{person}{Anne-Marie Brouwer}, \bibinfo{person}{Maarten~A Hogervorst}, \bibinfo{person}{Bob Oudejans}, \bibinfo{person}{Anthony~J Ries}, {and} \bibinfo{person}{Jonathan Touryan}.} \bibinfo{year}{2017}\natexlab{}.
\newblock \showarticletitle{EEG and eye tracking signatures of target encoding during structured visual search}.
\newblock \bibinfo{journal}{\emph{Frontiers in human neuroscience}}  \bibinfo{volume}{11} (\bibinfo{year}{2017}), \bibinfo{pages}{264}.
\newblock


\bibitem[Bylinskii et~al\mbox{.}(2018)]%
        {bylinskii2018different}
\bibfield{author}{\bibinfo{person}{Zoya Bylinskii}, \bibinfo{person}{Tilke Judd}, \bibinfo{person}{Aude Oliva}, \bibinfo{person}{Antonio Torralba}, {and} \bibinfo{person}{Fr{\'e}do Durand}.} \bibinfo{year}{2018}\natexlab{}.
\newblock \showarticletitle{What do different evaluation metrics tell us about saliency models?}
\newblock \bibinfo{journal}{\emph{IEEE transactions on pattern analysis and machine intelligence}} \bibinfo{volume}{41}, \bibinfo{number}{3} (\bibinfo{year}{2018}), \bibinfo{pages}{740--757}.
\newblock


\bibitem[Castelhano and Henderson(2007)]%
        {RN987}
\bibfield{author}{\bibinfo{person}{Monica~S. Castelhano} {and} \bibinfo{person}{John~M. Henderson}.} \bibinfo{year}{2007}\natexlab{}.
\newblock \showarticletitle{Initial scene representations facilitate eye movement guidance in visual search}.
\newblock \bibinfo{journal}{\emph{Journal of Experimental Psychology: Human Perception and Performance}} \bibinfo{volume}{33}, \bibinfo{number}{4} (\bibinfo{year}{2007}), \bibinfo{pages}{753--63}.
\newblock
\showISSN{0096-1523 (Print) 0096-1523 (Linking)}
\href{https://doi.org/10.1037/0096-1523.33.4.753}{doi:\nolinkurl{10.1037/0096-1523.33.4.753}}


\bibitem[Chiossi et~al\mbox{.}(2026)]%
        {chiossi2026anticipation}
\bibfield{author}{\bibinfo{person}{Francesco Chiossi}, \bibinfo{person}{Elnur Imamaliyev}, \bibinfo{person}{Martin Bleichner}, {and} \bibinfo{person}{Sven Mayer}.} \bibinfo{year}{2026}\natexlab{}.
\newblock \showarticletitle{Anticipation before action: EEG-based implicit intent detection for adaptive gaze interaction in mixed reality}. In \bibinfo{booktitle}{\emph{Proceedings of the 2026 CHI Conference on Human Factors in Computing Systems}}. \bibinfo{pages}{1--24}.
\newblock


\bibitem[Delorme and Makeig(2004)]%
        {delorme2004eeglab}
\bibfield{author}{\bibinfo{person}{Arnaud Delorme} {and} \bibinfo{person}{Scott Makeig}.} \bibinfo{year}{2004}\natexlab{}.
\newblock \showarticletitle{EEGLAB: an open source toolbox for analysis of single-trial EEG dynamics including independent component analysis}.
\newblock \bibinfo{journal}{\emph{Journal of Neuroscience Methods}} \bibinfo{volume}{134}, \bibinfo{number}{1} (\bibinfo{year}{2004}), \bibinfo{pages}{9--21}.
\newblock
\href{https://doi.org/10.1016/j.jneumeth.2003.10.009}{doi:\nolinkurl{10.1016/j.jneumeth.2003.10.009}}


\bibitem[Dolci et~al\mbox{.}(2023)]%
        {Dolci2023}
\bibfield{author}{\bibinfo{person}{Carola Dolci}, \bibinfo{person}{C.~Nico Boehler}, \bibinfo{person}{Elisa Santandrea}, \bibinfo{person}{Anneleen Dewulf}, \bibinfo{person}{Suliann Ben-Hamed}, \bibinfo{person}{Emiliano Macaluso}, \bibinfo{person}{Leonardo Chelazzi}, {and} \bibinfo{person}{Einat Rashal}.} \bibinfo{year}{2023}\natexlab{}.
\newblock \showarticletitle{Integrated effects of top-down attention and statistical learning during visual search: An EEG study}.
\newblock \bibinfo{journal}{\emph{Attention, Perception, \& Psychophysics}} \bibinfo{volume}{85}, \bibinfo{number}{6} (\bibinfo{year}{2023}), \bibinfo{pages}{1819--1833}.
\newblock
\href{https://doi.org/10.3758/s13414-023-02728-y}{doi:\nolinkurl{10.3758/s13414-023-02728-y}}


\bibitem[Droste et~al\mbox{.}(2020)]%
        {droste2020unified}
\bibfield{author}{\bibinfo{person}{Richard Droste}, \bibinfo{person}{Jianbo Jiao}, {and} \bibinfo{person}{J~Alison Noble}.} \bibinfo{year}{2020}\natexlab{}.
\newblock \showarticletitle{Unified image and video saliency modeling}. In \bibinfo{booktitle}{\emph{European conference on computer vision}}. Springer, \bibinfo{pages}{419--435}.
\newblock


\bibitem[Everett and Byrne(2004)]%
        {10.1145/985692.985780}
\bibfield{author}{\bibinfo{person}{Sarah~P. Everett} {and} \bibinfo{person}{Michael~D. Byrne}.} \bibinfo{year}{2004}\natexlab{}.
\newblock \showarticletitle{Unintended effects: varying icon spacing changes users' visual search strategy}. In \bibinfo{booktitle}{\emph{Proceedings of the SIGCHI Conference on Human Factors in Computing Systems}} (Vienna, Austria) \emph{(\bibinfo{series}{CHI '04})}. \bibinfo{publisher}{Association for Computing Machinery}, \bibinfo{address}{New York, NY, USA}, \bibinfo{pages}{695–702}.
\newblock
\showISBNx{1581137028}
\href{https://doi.org/10.1145/985692.985780}{doi:\nolinkurl{10.1145/985692.985780}}


\bibitem[Faber et~al\mbox{.}(2012)]%
        {faber2012mental}
\bibfield{author}{\bibinfo{person}{L{\'e}on~G Faber}, \bibinfo{person}{Natasha~M Maurits}, {and} \bibinfo{person}{Monicque~M Lorist}.} \bibinfo{year}{2012}\natexlab{}.
\newblock \showarticletitle{Mental fatigue affects visual selective attention}.
\newblock \bibinfo{journal}{\emph{PloS one}} \bibinfo{volume}{7}, \bibinfo{number}{10} (\bibinfo{year}{2012}), \bibinfo{pages}{e48073}.
\newblock


\bibitem[Fan et~al\mbox{.}(2015)]%
        {fan2015eeg}
\bibfield{author}{\bibinfo{person}{Xiaoli Fan}, \bibinfo{person}{Qianxiang Zhou}, \bibinfo{person}{Zhongqi Liu}, {and} \bibinfo{person}{Fang Xie}.} \bibinfo{year}{2015}\natexlab{}.
\newblock \showarticletitle{Electroencephalogram assessment of mental fatigue in visual search}.
\newblock \bibinfo{journal}{\emph{Bio-Medical Materials and Engineering}}  \bibinfo{volume}{26 Suppl 1} (\bibinfo{year}{2015}), \bibinfo{pages}{S1455--S1463}.
\newblock
\href{https://doi.org/10.3233/BME-151444}{doi:\nolinkurl{10.3233/BME-151444}}


\bibitem[Farahat et~al\mbox{.}(2019)]%
        {farahat2019convolutional}
\bibfield{author}{\bibinfo{person}{Amr Farahat}, \bibinfo{person}{Christoph Reichert}, \bibinfo{person}{Catherine~M Sweeney-Reed}, {and} \bibinfo{person}{Hermann Hinrichs}.} \bibinfo{year}{2019}\natexlab{}.
\newblock \showarticletitle{Convolutional neural networks for decoding of covert attention focus and saliency maps for EEG feature visualization}.
\newblock \bibinfo{journal}{\emph{Journal of neural engineering}} \bibinfo{volume}{16}, \bibinfo{number}{6} (\bibinfo{year}{2019}), \bibinfo{pages}{066010}.
\newblock


\bibitem[Gramfort et~al\mbox{.}(2013)]%
        {GramfortEtAl2013a}
\bibfield{author}{\bibinfo{person}{Alexandre Gramfort}, \bibinfo{person}{Martin Luessi}, \bibinfo{person}{Eric Larson}, \bibinfo{person}{Denis~A. Engemann}, \bibinfo{person}{Daniel Strohmeier}, \bibinfo{person}{Christian Brodbeck}, \bibinfo{person}{Roman Goj}, \bibinfo{person}{Mainak Jas}, \bibinfo{person}{Teon Brooks}, \bibinfo{person}{Lauri Parkkonen}, {and} \bibinfo{person}{Matti~S. H{\"a}m{\"a}l{\"a}inen}.} \bibinfo{year}{2013}\natexlab{}.
\newblock \showarticletitle{{{MEG}} and {{EEG}} Data Analysis with {{MNE}}-{{Python}}}.
\newblock \bibinfo{journal}{\emph{Frontiers in Neuroscience}} \bibinfo{volume}{7}, \bibinfo{number}{267} (\bibinfo{year}{2013}), \bibinfo{pages}{1--13}.
\newblock
\href{https://doi.org/10.3389/fnins.2013.00267}{doi:\nolinkurl{10.3389/fnins.2013.00267}}


\bibitem[Haas(2014)]%
        {unity}
\bibfield{author}{\bibinfo{person}{John~K Haas}.} \bibinfo{year}{2014}\natexlab{}.
\newblock \showarticletitle{A History of the Unity Game Engine}.
\newblock
\urldef\tempurl%
\url{https://api.semanticscholar.org/CorpusID:86824974}
\showURL{%
\tempurl}


\bibitem[Hadnett-Hunter et~al\mbox{.}(2019)]%
        {hadnett2019effect}
\bibfield{author}{\bibinfo{person}{Jacob Hadnett-Hunter}, \bibinfo{person}{George Nicolaou}, \bibinfo{person}{Eamonn O'Neill}, {and} \bibinfo{person}{Michael Proulx}.} \bibinfo{year}{2019}\natexlab{}.
\newblock \showarticletitle{The effect of task on visual attention in interactive virtual environments}.
\newblock \bibinfo{journal}{\emph{ACM Transactions on Applied Perception (TAP)}} \bibinfo{volume}{16}, \bibinfo{number}{3} (\bibinfo{year}{2019}), \bibinfo{pages}{1--17}.
\newblock


\bibitem[Hancock et~al\mbox{.}(2013)]%
        {hancock2013improving}
\bibfield{author}{\bibinfo{person}{Peter~A Hancock}, \bibinfo{person}{Joseph~E Mercado}, \bibinfo{person}{James Merlo}, {and} \bibinfo{person}{Jan~BF Van~Erp}.} \bibinfo{year}{2013}\natexlab{}.
\newblock \showarticletitle{Improving target detection in visual search through the augmenting multi-sensory cues}.
\newblock \bibinfo{journal}{\emph{Ergonomics}} \bibinfo{volume}{56}, \bibinfo{number}{5} (\bibinfo{year}{2013}), \bibinfo{pages}{729--738}.
\newblock


\bibitem[Hopf et~al\mbox{.}(2000)]%
        {10.1093/cercor/10.12.1233}
\bibfield{author}{\bibinfo{person}{Jens-Max Hopf}, \bibinfo{person}{Steven~J. Luck}, \bibinfo{person}{Massimo Girelli}, \bibinfo{person}{Tilman Hagner}, \bibinfo{person}{George~R. Mangun}, \bibinfo{person}{Henning Scheich}, {and} \bibinfo{person}{Hans-Jochen Heinze}.} \bibinfo{year}{2000}\natexlab{}.
\newblock \showarticletitle{Neural Sources of Focused Attention in Visual Search}.
\newblock \bibinfo{journal}{\emph{Cerebral Cortex}} \bibinfo{volume}{10}, \bibinfo{number}{12} (\bibinfo{date}{12} \bibinfo{year}{2000}), \bibinfo{pages}{1233--1241}.
\newblock
\showISSN{1047-3211}
\showeprint{https://academic.oup.com/cercor/article-pdf/10/12/1233/9751024/1001233.pdf}
\href{https://doi.org/10.1093/cercor/10.12.1233}{doi:\nolinkurl{10.1093/cercor/10.12.1233}}


\bibitem[Hosseini et~al\mbox{.}(2025)]%
        {hosseini2025sum}
\bibfield{author}{\bibinfo{person}{Alireza Hosseini}, \bibinfo{person}{Amirhossein Kazerouni}, \bibinfo{person}{Saeed Akhavan}, \bibinfo{person}{Michael Brudno}, {and} \bibinfo{person}{Babak Taati}.} \bibinfo{year}{2025}\natexlab{}.
\newblock \showarticletitle{Sum: Saliency unification through mamba for visual attention modeling}. In \bibinfo{booktitle}{\emph{2025 IEEE/CVF Winter Conference on Applications of Computer Vision (WACV)}}. IEEE, \bibinfo{pages}{1597--1607}.
\newblock


\bibitem[Itti et~al\mbox{.}(2002)]%
        {itti2002model}
\bibfield{author}{\bibinfo{person}{Laurent Itti}, \bibinfo{person}{Christof Koch}, {and} \bibinfo{person}{Ernst Niebur}.} \bibinfo{year}{2002}\natexlab{}.
\newblock \showarticletitle{A model of saliency-based visual attention for rapid scene analysis}.
\newblock \bibinfo{journal}{\emph{IEEE Transactions on pattern analysis and machine intelligence}} \bibinfo{volume}{20}, \bibinfo{number}{11} (\bibinfo{year}{2002}), \bibinfo{pages}{1254--1259}.
\newblock


\bibitem[Kamienkowski et~al\mbox{.}(2012)]%
        {kamienkowski2012fixation}
\bibfield{author}{\bibinfo{person}{Juan~E Kamienkowski}, \bibinfo{person}{Matias~J Ison}, \bibinfo{person}{Rodrigo~Quian Quiroga}, {and} \bibinfo{person}{Mariano Sigman}.} \bibinfo{year}{2012}\natexlab{}.
\newblock \showarticletitle{Fixation-related potentials in visual search: A combined EEG and eye tracking study}.
\newblock \bibinfo{journal}{\emph{Journal of vision}} \bibinfo{volume}{12}, \bibinfo{number}{7} (\bibinfo{year}{2012}), \bibinfo{pages}{4--4}.
\newblock


\bibitem[Kastrati et~al\mbox{.}(2021)]%
        {kastrati2021eegeyenet}
\bibfield{author}{\bibinfo{person}{Ard Kastrati}, \bibinfo{person}{Martyna~Beata P{\l}omecka}, \bibinfo{person}{Dami{\'a}n Pascual}, \bibinfo{person}{Lukas Wolf}, \bibinfo{person}{Victor Gillioz}, \bibinfo{person}{Roger Wattenhofer}, {and} \bibinfo{person}{Nicolas Langer}.} \bibinfo{year}{2021}\natexlab{}.
\newblock \showarticletitle{EEGEyeNet: a simultaneous electroencephalography and eye-tracking dataset and benchmark for eye movement prediction}.
\newblock \bibinfo{journal}{\emph{arXiv preprint arXiv:2111.05100}} (\bibinfo{year}{2021}).
\newblock


\bibitem[Kaunitz et~al\mbox{.}(2014)]%
        {kaunitz2014looking}
\bibfield{author}{\bibinfo{person}{Lisandro~N Kaunitz}, \bibinfo{person}{Juan~E Kamienkowski}, \bibinfo{person}{Alexander Varatharajah}, \bibinfo{person}{Mariano Sigman}, \bibinfo{person}{Rodrigo~Quian Quiroga}, {and} \bibinfo{person}{Matias~J Ison}.} \bibinfo{year}{2014}\natexlab{}.
\newblock \showarticletitle{Looking for a face in the crowd: Fixation-related potentials in an eye-movement visual search task}.
\newblock \bibinfo{journal}{\emph{NeuroImage}}  \bibinfo{volume}{89} (\bibinfo{year}{2014}), \bibinfo{pages}{297--305}.
\newblock


\bibitem[Khaleghi et~al\mbox{.}(2022)]%
        {khaleghi2022visual}
\bibfield{author}{\bibinfo{person}{Nastaran Khaleghi}, \bibinfo{person}{Tohid~Yousefi Rezaii}, \bibinfo{person}{Soosan Beheshti}, \bibinfo{person}{Saeed Meshgini}, \bibinfo{person}{Sobhan Sheykhivand}, {and} \bibinfo{person}{Sebelan Danishvar}.} \bibinfo{year}{2022}\natexlab{}.
\newblock \showarticletitle{Visual saliency and image reconstruction from EEG signals via an effective geometric deep network-based generative adversarial network}.
\newblock \bibinfo{journal}{\emph{Electronics}} \bibinfo{volume}{11}, \bibinfo{number}{21} (\bibinfo{year}{2022}), \bibinfo{pages}{3637}.
\newblock


\bibitem[Koch and Ullman(1987)]%
        {koch1987shifts}
\bibfield{author}{\bibinfo{person}{Christof Koch} {and} \bibinfo{person}{Shimon Ullman}.} \bibinfo{year}{1987}\natexlab{}.
\newblock \showarticletitle{Shifts in selective visual attention: towards the underlying neural circuitry}.
\newblock In \bibinfo{booktitle}{\emph{Matters of intelligence: Conceptual structures in cognitive neuroscience}}. \bibinfo{publisher}{Springer}, \bibinfo{pages}{115--141}.
\newblock


\bibitem[Kruthiventi et~al\mbox{.}(2017)]%
        {kruthiventi2017deepfix}
\bibfield{author}{\bibinfo{person}{Srinivas~SS Kruthiventi}, \bibinfo{person}{Kumar Ayush}, {and} \bibinfo{person}{R~Venkatesh Babu}.} \bibinfo{year}{2017}\natexlab{}.
\newblock \showarticletitle{Deepfix: A fully convolutional neural network for predicting human eye fixations}.
\newblock \bibinfo{journal}{\emph{IEEE Transactions on Image Processing}} \bibinfo{volume}{26}, \bibinfo{number}{9} (\bibinfo{year}{2017}), \bibinfo{pages}{4446--4456}.
\newblock


\bibitem[K{\"u}mmerer et~al\mbox{.}({[n.\,d.]})]%
        {mit-tuebingen-saliency-benchmark}
\bibfield{author}{\bibinfo{person}{Matthias K{\"u}mmerer}, \bibinfo{person}{Zoya Bylinskii}, \bibinfo{person}{Tilke Judd}, \bibinfo{person}{Ali Borji}, \bibinfo{person}{Laurent Itti}, \bibinfo{person}{Fr{\'e}do Durand}, \bibinfo{person}{Aude Oliva}, \bibinfo{person}{Antonio Torralba}, {and} \bibinfo{person}{Matthias Bethge}.} \bibinfo{year}{[n.\,d.]}\natexlab{}.
\newblock \bibinfo{title}{MIT/Tübingen Saliency Benchmark}.
\newblock \bibinfo{howpublished}{https://saliency.tuebingen.ai/}.
\newblock


\bibitem[K{\"u}mmerer et~al\mbox{.}(2014)]%
        {kummerer2014deep}
\bibfield{author}{\bibinfo{person}{Matthias K{\"u}mmerer}, \bibinfo{person}{Lucas Theis}, {and} \bibinfo{person}{Matthias Bethge}.} \bibinfo{year}{2014}\natexlab{}.
\newblock \showarticletitle{Deep gaze i: Boosting saliency prediction with feature maps trained on imagenet}.
\newblock \bibinfo{journal}{\emph{arXiv preprint arXiv:1411.1045}} (\bibinfo{year}{2014}).
\newblock


\bibitem[K{\"u}mmerer et~al\mbox{.}(2016)]%
        {kummerer2016deepgaze}
\bibfield{author}{\bibinfo{person}{Matthias K{\"u}mmerer}, \bibinfo{person}{Thomas~SA Wallis}, {and} \bibinfo{person}{Matthias Bethge}.} \bibinfo{year}{2016}\natexlab{}.
\newblock \showarticletitle{DeepGaze II: Reading fixations from deep features trained on object recognition}.
\newblock \bibinfo{journal}{\emph{arXiv preprint arXiv:1610.01563}} (\bibinfo{year}{2016}).
\newblock


\bibitem[Lee et~al\mbox{.}(2007)]%
        {lee2007visual}
\bibfield{author}{\bibinfo{person}{Yi-Ching Lee}, \bibinfo{person}{John~D Lee}, {and} \bibinfo{person}{Linda Ng~Boyle}.} \bibinfo{year}{2007}\natexlab{}.
\newblock \showarticletitle{Visual attention in driving: The effects of cognitive load and visual disruption}.
\newblock \bibinfo{journal}{\emph{Human Factors}} \bibinfo{volume}{49}, \bibinfo{number}{4} (\bibinfo{year}{2007}), \bibinfo{pages}{721--733}.
\newblock


\bibitem[Liesefeld et~al\mbox{.}(2017)]%
        {LIESEFELD2017166}
\bibfield{author}{\bibinfo{person}{Heinrich~René Liesefeld}, \bibinfo{person}{Anna~Marie Liesefeld}, \bibinfo{person}{Thomas Töllner}, {and} \bibinfo{person}{Hermann~J. Müller}.} \bibinfo{year}{2017}\natexlab{}.
\newblock \showarticletitle{Attentional capture in visual search: Capture and post-capture dynamics revealed by EEG}.
\newblock \bibinfo{journal}{\emph{NeuroImage}}  \bibinfo{volume}{156} (\bibinfo{year}{2017}), \bibinfo{pages}{166--173}.
\newblock
\showISSN{1053-8119}
\href{https://doi.org/10.1016/j.neuroimage.2017.05.016}{doi:\nolinkurl{10.1016/j.neuroimage.2017.05.016}}


\bibitem[Liu et~al\mbox{.}(2021)]%
        {liu2021visual}
\bibfield{author}{\bibinfo{person}{Nian Liu}, \bibinfo{person}{Ni Zhang}, \bibinfo{person}{Kaiyuan Wan}, \bibinfo{person}{Ling Shao}, {and} \bibinfo{person}{Junwei Han}.} \bibinfo{year}{2021}\natexlab{}.
\newblock \showarticletitle{Visual saliency transformer}. In \bibinfo{booktitle}{\emph{Proceedings of the IEEE/CVF international conference on computer vision}}. \bibinfo{pages}{4722--4732}.
\newblock


\bibitem[Lou et~al\mbox{.}(2022)]%
        {lou2022transalnet}
\bibfield{author}{\bibinfo{person}{Jianxun Lou}, \bibinfo{person}{Hanhe Lin}, \bibinfo{person}{David Marshall}, \bibinfo{person}{Dietmar Saupe}, {and} \bibinfo{person}{Hantao Liu}.} \bibinfo{year}{2022}\natexlab{}.
\newblock \showarticletitle{TranSalNet: Towards perceptually relevant visual saliency prediction}.
\newblock \bibinfo{journal}{\emph{Neurocomputing}}  \bibinfo{volume}{494} (\bibinfo{year}{2022}), \bibinfo{pages}{455--467}.
\newblock


\bibitem[Ludwig et~al\mbox{.}(2009)]%
        {ludwig2009using}
\bibfield{author}{\bibinfo{person}{Kip Ludwig}, \bibinfo{person}{Rachel Miriani}, \bibinfo{person}{Nicholas Langhals}, \bibinfo{person}{Michael Joseph}, \bibinfo{person}{David Anderson}, {and} \bibinfo{person}{Daryl Kipke}.} \bibinfo{year}{2009}\natexlab{}.
\newblock \showarticletitle{Using a Common Average Reference to Improve Cortical Neuron Recordings From Microelectrode Arrays}.
\newblock \bibinfo{journal}{\emph{Journal of neurophysiology}}  \bibinfo{volume}{101} (\bibinfo{date}{03} \bibinfo{year}{2009}), \bibinfo{pages}{1679--89}.
\newblock
\href{https://doi.org/10.1152/jn.90989.2008}{doi:\nolinkurl{10.1152/jn.90989.2008}}


\bibitem[M{\"u}ller et~al\mbox{.}(2022)]%
        {muller2022designing}
\bibfield{author}{\bibinfo{person}{Philipp M{\"u}ller}, \bibinfo{person}{Sander Staal}, \bibinfo{person}{Mihai B{\^a}ce}, {and} \bibinfo{person}{Andreas Bulling}.} \bibinfo{year}{2022}\natexlab{}.
\newblock \showarticletitle{Designing for noticeability: Understanding the impact of visual importance on desktop notifications}. In \bibinfo{booktitle}{\emph{Proceedings of the 2022 CHI Conference on Human Factors in Computing Systems}}. \bibinfo{pages}{1--13}.
\newblock


\bibitem[Olsen(2012)]%
        {olsen2012tobii}
\bibfield{author}{\bibinfo{person}{Anneli Olsen}.} \bibinfo{year}{2012}\natexlab{}.
\newblock \showarticletitle{The Tobii IVT Fixation Filter Algorithm description}.
\newblock
\urldef\tempurl%
\url{https://api.semanticscholar.org/CorpusID:52834703}
\showURL{%
\tempurl}


\bibitem[Ramanishka et~al\mbox{.}(2017)]%
        {ramanishka2017top}
\bibfield{author}{\bibinfo{person}{Vasili Ramanishka}, \bibinfo{person}{Abir Das}, \bibinfo{person}{Jianming Zhang}, {and} \bibinfo{person}{Kate Saenko}.} \bibinfo{year}{2017}\natexlab{}.
\newblock \showarticletitle{Top-down visual saliency guided by captions}. In \bibinfo{booktitle}{\emph{Proceedings of the IEEE conference on computer vision and pattern recognition}}. \bibinfo{pages}{7206--7215}.
\newblock


\bibitem[Samiei and Clark(2022)]%
        {samiei2022predictingvisualattentiondistraction}
\bibfield{author}{\bibinfo{person}{Manoosh Samiei} {and} \bibinfo{person}{James~J. Clark}.} \bibinfo{year}{2022}\natexlab{}.
\newblock \bibinfo{title}{Predicting Visual Attention and Distraction During Visual Search Using Convolutional Neural Networks}.
\newblock
\showeprint[arxiv]{2210.15093}~[cs.CV]
\urldef\tempurl%
\url{https://arxiv.org/abs/2210.15093}
\showURL{%
\tempurl}


\bibitem[Sharma(2021)]%
        {sharma2021mental}
\bibfield{author}{\bibinfo{person}{Mansi Sharma}.} \bibinfo{year}{2021}\natexlab{}.
\newblock \showarticletitle{Mental workload classification}.
\newblock \bibinfo{journal}{\emph{Poster.[Google Scholar]}} (\bibinfo{year}{2021}).
\newblock


\bibitem[Sharma et~al\mbox{.}(2023)]%
        {sharma2023implicit}
\bibfield{author}{\bibinfo{person}{Mansi Sharma}, \bibinfo{person}{Shuang Chen}, \bibinfo{person}{Philipp M{\"u}ller}, \bibinfo{person}{Maurice Rekrut}, {and} \bibinfo{person}{Antonio Kr{\"u}ger}.} \bibinfo{year}{2023}\natexlab{}.
\newblock \showarticletitle{Implicit Search Intent Recognition using EEG and Eye Tracking: Novel Dataset and Cross-User Prediction}. In \bibinfo{booktitle}{\emph{Proceedings of the 25th International Conference on Multimodal Interaction}}. \bibinfo{pages}{345--354}.
\newblock


\bibitem[Sharma et~al\mbox{.}(2024)]%
        {sharma2024distinguishing}
\bibfield{author}{\bibinfo{person}{Mansi Sharma}, \bibinfo{person}{Camilo~Andr{\'e}s Mart{\'\i}nez~Mart{\'\i}nez}, \bibinfo{person}{Benedikt~Emanuel Wirth}, \bibinfo{person}{Antonio Kr{\"u}ger}, {and} \bibinfo{person}{Philipp M{\"u}ller}.} \bibinfo{year}{2024}\natexlab{}.
\newblock \showarticletitle{Distinguishing target and non-target fixations with eeg and eye tracking in realistic visual scenes}. In \bibinfo{booktitle}{\emph{Proceedings of the 26th International Conference on Multimodal Interaction}}. \bibinfo{pages}{459--468}.
\newblock


\bibitem[Sharma et~al\mbox{.}(2022)]%
        {sharma2022towards}
\bibfield{author}{\bibinfo{person}{Mansi Sharma}, \bibinfo{person}{Maurice Rekrut}, \bibinfo{person}{Jan Alexandersson}, {and} \bibinfo{person}{Antonio Kr{\"u}ger}.} \bibinfo{year}{2022}\natexlab{}.
\newblock \showarticletitle{Towards improving eeg-based intent recognition in visual search tasks}. In \bibinfo{booktitle}{\emph{International Conference on Neural Information Processing}}. Springer, \bibinfo{pages}{604--615}.
\newblock


\bibitem[Strohm et~al\mbox{.}(2024)]%
        {10.1145/3655603}
\bibfield{author}{\bibinfo{person}{Florian Strohm}, \bibinfo{person}{Mihai B\^{a}ce}, {and} \bibinfo{person}{Andreas Bulling}.} \bibinfo{year}{2024}\natexlab{}.
\newblock \showarticletitle{Learning User Embeddings from Human Gaze for Personalised Saliency Prediction}.
\newblock \bibinfo{journal}{\emph{Proc. ACM Hum.-Comput. Interact.}} \bibinfo{volume}{8}, \bibinfo{number}{ETRA}, Article \bibinfo{articleno}{229} (\bibinfo{date}{May} \bibinfo{year}{2024}), \bibinfo{numpages}{16}~pages.
\newblock
\href{https://doi.org/10.1145/3655603}{doi:\nolinkurl{10.1145/3655603}}


\bibitem[Trabulsi et~al\mbox{.}(2021)]%
        {trabulsi2021optimizing}
\bibfield{author}{\bibinfo{person}{Julia Trabulsi}, \bibinfo{person}{Kian Norouzi}, \bibinfo{person}{Seidi Suurmets}, \bibinfo{person}{Mike Storm}, {and} \bibinfo{person}{Thomas~Zo{\"e}ga Rams{\o}y}.} \bibinfo{year}{2021}\natexlab{}.
\newblock \showarticletitle{Optimizing fixation filters for eye-tracking on small screens}.
\newblock \bibinfo{journal}{\emph{Frontiers in neuroscience}} (\bibinfo{year}{2021}), \bibinfo{pages}{1257}.
\newblock
\href{https://doi.org/10.3389/fnins.2021.578439}{doi:\nolinkurl{10.3389/fnins.2021.578439}}


\bibitem[Van~Humbeeck et~al\mbox{.}(2018)]%
        {van2018presaccadic}
\bibfield{author}{\bibinfo{person}{Nathalie Van~Humbeeck}, \bibinfo{person}{Radha~Nila Meghanathan}, \bibinfo{person}{Johan Wagemans}, \bibinfo{person}{Cees Van~Leeuwen}, {and} \bibinfo{person}{Andrey~R Nikolaev}.} \bibinfo{year}{2018}\natexlab{}.
\newblock \showarticletitle{Presaccadic EEG activity predicts visual saliency in free-viewing contour integration}.
\newblock \bibinfo{journal}{\emph{Psychophysiology}} \bibinfo{volume}{55}, \bibinfo{number}{12} (\bibinfo{year}{2018}), \bibinfo{pages}{e13267}.
\newblock


\bibitem[Vig et~al\mbox{.}(2014)]%
        {vig2014large}
\bibfield{author}{\bibinfo{person}{Eleonora Vig}, \bibinfo{person}{Michael Dorr}, {and} \bibinfo{person}{David Cox}.} \bibinfo{year}{2014}\natexlab{}.
\newblock \showarticletitle{Large-scale optimization of hierarchical features for saliency prediction in natural images}. In \bibinfo{booktitle}{\emph{Proceedings of the IEEE conference on computer vision and pattern recognition}}. \bibinfo{pages}{2798--2805}.
\newblock


\bibitem[Vortmann et~al\mbox{.}(2022)]%
        {vortmann2022multimodal}
\bibfield{author}{\bibinfo{person}{Lisa-Marie Vortmann}, \bibinfo{person}{Simon Ceh}, {and} \bibinfo{person}{Felix Putze}.} \bibinfo{year}{2022}\natexlab{}.
\newblock \showarticletitle{Multimodal eeg and eye tracking feature fusion approaches for attention classification in hybrid bcis}.
\newblock \bibinfo{journal}{\emph{Frontiers in Computer Science}}  \bibinfo{volume}{4} (\bibinfo{year}{2022}), \bibinfo{pages}{780580}.
\newblock


\bibitem[Vozniak et~al\mbox{.}(2023)]%
        {vozniak2023context}
\bibfield{author}{\bibinfo{person}{Igor Vozniak}, \bibinfo{person}{Philipp M{\"u}ller}, \bibinfo{person}{Lorena Hell}, \bibinfo{person}{Nils Lipp}, \bibinfo{person}{Ahmed Abouelazm}, {and} \bibinfo{person}{Christian M{\"u}ller}.} \bibinfo{year}{2023}\natexlab{}.
\newblock \showarticletitle{Context-empowered visual attention prediction in pedestrian scenarios}. In \bibinfo{booktitle}{\emph{Proceedings of the IEEE/CVF Winter Conference on Applications of Computer Vision}}. \bibinfo{pages}{950--960}.
\newblock


\bibitem[Wang et~al\mbox{.}(2024)]%
        {wang2024salchartqa}
\bibfield{author}{\bibinfo{person}{Yao Wang}, \bibinfo{person}{Weitian Wang}, \bibinfo{person}{Abdullah Abdelhafez}, \bibinfo{person}{Mayar Elfares}, \bibinfo{person}{Zhiming Hu}, \bibinfo{person}{Mihai B{\^a}ce}, {and} \bibinfo{person}{Andreas Bulling}.} \bibinfo{year}{2024}\natexlab{}.
\newblock \showarticletitle{SalChartQA: Question-driven saliency on information visualisations}. In \bibinfo{booktitle}{\emph{Proceedings of the 2024 CHI Conference on Human Factors in Computing Systems}}. \bibinfo{pages}{1--14}.
\newblock


\bibitem[Wolfe and Horowitz(2017)]%
        {RN986}
\bibfield{author}{\bibinfo{person}{Jeremy~M. Wolfe} {and} \bibinfo{person}{Todd~S. Horowitz}.} \bibinfo{year}{2017}\natexlab{}.
\newblock \showarticletitle{Five factors that guide attention in visual search}.
\newblock \bibinfo{journal}{\emph{Nature Human Behaviour}} \bibinfo{volume}{1}, \bibinfo{number}{3} (\bibinfo{year}{2017}).
\newblock
\showISSN{2397-3374 (Electronic) 2397-3374 (Linking)}
\href{https://doi.org/10.1038/s41562-017-0058}{doi:\nolinkurl{10.1038/s41562-017-0058}}


\bibitem[Wolfe et~al\mbox{.}(2011)]%
        {RN575}
\bibfield{author}{\bibinfo{person}{Jeremy~M. Wolfe}, \bibinfo{person}{Melissa~L. Vo}, \bibinfo{person}{Karla~K. Evans}, {and} \bibinfo{person}{Michelle~R. Greene}.} \bibinfo{year}{2011}\natexlab{}.
\newblock \showarticletitle{Visual search in scenes involves selective and nonselective pathways}.
\newblock \bibinfo{journal}{\emph{Trends in Cognitive Sciences}} \bibinfo{volume}{15}, \bibinfo{number}{2} (\bibinfo{year}{2011}), \bibinfo{pages}{77--84}.
\newblock
\showISSN{1879-307X (Electronic) 1364-6613 (Linking)}
\href{https://doi.org/10.1016/j.tics.2010.12.001}{doi:\nolinkurl{10.1016/j.tics.2010.12.001}}


\bibitem[Xu et~al\mbox{.}(2018)]%
        {xu2018personalized}
\bibfield{author}{\bibinfo{person}{Yanyu Xu}, \bibinfo{person}{Shenghua Gao}, \bibinfo{person}{Junru Wu}, \bibinfo{person}{Nianyi Li}, {and} \bibinfo{person}{Jingyi Yu}.} \bibinfo{year}{2018}\natexlab{}.
\newblock \showarticletitle{Personalized saliency and its prediction}.
\newblock \bibinfo{journal}{\emph{IEEE transactions on pattern analysis and machine intelligence}} \bibinfo{volume}{41}, \bibinfo{number}{12} (\bibinfo{year}{2018}), \bibinfo{pages}{2975--2989}.
\newblock


\bibitem[Xue et~al\mbox{.}(2025)]%
        {xue2025few}
\bibfield{author}{\bibinfo{person}{Ruoyu Xue}, \bibinfo{person}{Jingyi Xu}, \bibinfo{person}{Sounak Mondal}, \bibinfo{person}{Hieu Le}, \bibinfo{person}{Greg Zelinsky}, \bibinfo{person}{Minh Hoai}, {and} \bibinfo{person}{Dimitris Samaras}.} \bibinfo{year}{2025}\natexlab{}.
\newblock \showarticletitle{Few-shot personalized scanpath prediction}. In \bibinfo{booktitle}{\emph{Proceedings of the IEEE/CVF Conference on Computer Vision and Pattern Recognition}}. \bibinfo{pages}{13497--13507}.
\newblock


\bibitem[Zagermann et~al\mbox{.}(2016)]%
        {zagermann2016measuring}
\bibfield{author}{\bibinfo{person}{Johannes Zagermann}, \bibinfo{person}{Ulrike Pfeil}, {and} \bibinfo{person}{Harald Reiterer}.} \bibinfo{year}{2016}\natexlab{}.
\newblock \showarticletitle{Measuring cognitive load using eye tracking technology in visual computing}. In \bibinfo{booktitle}{\emph{Proceedings of the sixth workshop on beyond time and errors on novel evaluation methods for visualization}}. \bibinfo{pages}{78--85}.
\newblock


\bibitem[Zheng et~al\mbox{.}(2018)]%
        {zheng2018task}
\bibfield{author}{\bibinfo{person}{Quanlong Zheng}, \bibinfo{person}{Jianbo Jiao}, \bibinfo{person}{Ying Cao}, {and} \bibinfo{person}{Rynson~WH Lau}.} \bibinfo{year}{2018}\natexlab{}.
\newblock \showarticletitle{Task-driven webpage saliency}. In \bibinfo{booktitle}{\emph{Proceedings of the European conference on computer vision (ECCV)}}. \bibinfo{pages}{287--302}.
\newblock


\end{thebibliography}

\end{document}